\documentclass[preprint,3p,twocolumn,sort&compress]{elsarticle}

\usepackage{graphicx,ulem}
\usepackage{longtable}
\usepackage{float}
\usepackage{dcolumn}
\usepackage{graphics,epsfig}
\usepackage{amsmath,amssymb,latexsym,mathrsfs}
\usepackage{bm}
\usepackage{color}
\usepackage{subfigure}
\usepackage{multirow}
\usepackage{url}

\newcommand{\eV}{\,\mathrm{eV}}

\newcommand{\mlight}{m_{\mathrm{light}}}

\newcommand{\lcdm}{\Lambda\mathrm{CDM}}
\newcommand{\obh}{\Omega_b h^2}
\newcommand{\och}{\Omega_c h^2}
\newcommand{\loga}{\ln[10^{10}A_s]}
\newcommand{\htype}{h_\mathrm{type}}

\DeclareMathOperator{\sgn}{sgn}

\def\aap{\ref@jnl{A\&A}}                

\journal{}

\begin{document}

\begin{frontmatter}

\title{A novel approach to quantifying the sensitivity of current and future cosmological datasets to the neutrino mass ordering through Bayesian hierarchical modeling}

\author[okc]{Martina Gerbino\corref{cor1}}
\ead{martina.gerbino@fysik.su.se}	

\author[fe]{Massimiliano Lattanzi\corref{cor1}}
\ead{lattanzi@fe.infn.it}

\author[ific]{Olga Mena}

\author[okc,nordita,umich]{Katherine Freese}

\cortext[cor1]{Corresponding author}

\address[okc]{The Oskar Klein Centre for Cosmoparticle Physics, Department of Physics, Stockholm University, AlbaNova, SE-106 91 Stockholm, Sweden}
\address[fe]{Dipartimento di Fisica e Scienze della Terra, Universit\`a di Ferrara and INFN, Sezione di Ferrara, Polo Scientifico e Tecnologico - Edificio C Via Saragat, 1, I-44122 Ferrara, Italy}
\address[ific]{IFIC, Universidad de Valencia-CSIC, 46071, Valencia, Spain}
\address[nordita]{The Nordic Institute for Theoretical Physics (NORDITA), Roslagstullsbacken 23, SE-106 91 Stockholm, Sweden}
\address[umich]{Michigan Center for Theoretical Physics, Department of Physics, University of Michigan, Ann Arbor, MI 48109, USA}


\begin{abstract}
We present a novel approach to derive constraints on neutrino masses, as well as on other cosmological parameters, from cosmological data,
while taking into account our ignorance of the neutrino mass ordering.
We derive constraints from a combination of current as well as future cosmological datasets on the total neutrino mass $M_\nu$ and on the
mass fractions $f_{\nu,i}=m_i/M_\nu$ (where the index $i=1,2,3$ indicates the three mass eigenstates) carried by each of the mass eigenstates $m_i$, after
marginalizing over the (unknown) neutrino mass ordering, either normal ordering (NH) or inverted ordering (IH). The bounds on all the
cosmological parameters, including those on the total neutrino mass,
take therefore into account the uncertainty related to our ignorance of the mass hierarchy that is actually realized in nature.
This novel approach is carried out in the framework of Bayesian analysis of a typical hierarchical problem, 
where the distribution of the parameters of the model depends on further parameters, the \textit{hyperparameters}. In this context,
the choice of the neutrino mass ordering is modeled via the discrete \textit{hyperparameter} $\htype$, which we introduce in the usual
Markov chain analysis. The preference from
cosmological data for either the NH or the IH 
scenarios is then simply encoded in the posterior distribution of the
\textit{hyperparameter} itself. Current cosmic microwave
background (CMB) measurements assign equal odds to the two hierarchies, and are thus unable to distinguish between them.
However, after the addition of baryon acoustic oscillation (BAO) measurements, a
weak preference for the normal hierarchical scenario appears, with odds of 4:3 from Planck temperature and large-scale
polarization in combination with BAO (3:2 if small-scale polarization is also included). 
Concerning next-generation cosmological experiments, forecasts suggest that the combination of
upcoming CMB (COrE) and BAO surveys (DESI) may determine the neutrino mass hierarchy
at a high statistical significance if the mass is very close to the minimal value allowed by oscillation experiments, as for NH and a fiducial value of 
$M_\nu=0.06\,\eV$ there is a 9:1 preference of normal versus inverted hierarchy. 
On the contrary, if the sum of the masses is of the order of $0.1\,\eV$ or larger, even future cosmological observations
will be inconclusive.  
The innovative statistical strategy exploited here represents a very simple,
efficient and robust tool to study the sensitivity of present and
future cosmological data to the neutrino mass hierarchy, and a sound competitor to the
standard Bayesian model comparison. The unbiased limit on
$M_\nu$ we obtain is crucial for ongoing and planned neutrinoless double beta decay searches.
\end{abstract}


\end{frontmatter}

\section{Introduction}\label{sec:intro}
According to the standard theory of neutrino oscillations (see e.g.~\cite{pdg2016} for an updated review and relevant references), the observed neutrino flavours $\nu_\alpha$ are a superposition of the massive eigenstates $\nu_i$:
\begin{equation}
|\nu_\alpha>=\sum_i U^{*}_{\alpha i} |\nu_i>
\label{eq:mix}
\end{equation}
where the index $\alpha$ can be any of the three active neutrino flavours $e,\mu,\tau$, the index $i=1,2,3$ runs over the three massive eigenstates and $U$ is the Pontecorvo-Maki-Nakagawa-Sakata mixing matrix, containing the neutrino mixing angles as well as the CP violating phases (one Dirac phase, 
as well as two additional Majorana phases, that are non vanishing only if neutrinos are Majorana particles). 

Cosmological measurements of the cosmic microwave background (hereafter CMB) anisotropies and of the spatial distribution of
galaxies provide the tightest bounds on the total neutrino mass,
defined as the sum of the three neutrino mass eigenstates, i.e. 
$M_\nu\equiv\sum m_{\nu,i}\equiv m_1+m_2+m_3$. The most reliable bound
that can be obtained combining Planck data with external datasets is $M_\nu
<0.21$~eV (at 95\% CL\footnote{We would like to warn the reader that the abbreviation ``CL'' is generally reserved for frequentist confidence level, whereas throughout this work we refer to bayesian credible intervals (see e.g. \cite{bayes}). In doing so, we have decided to adopt the common behaviour of speaking of confidence intervals even in the bayesian framework.})~\cite{Ade:2015xua}, from temperature plus large-scale polarization
CMB anisotropies and baryon acoustic oscillation (BAO) data (see also
Refs~\cite{Palanque-Delabrouille:2015pga,DiValentino:2015sam,
Cuesta:2015iho,Giusarma:2016phn,Hannestad:2016fog} for constraints obtained by the combination of additional datasets and/or in more extended cosmological scenarios). 
From neutrino oscillation data, we know that at least two out of the three mass
eigenstates should be massive, as two different
mass splittings are measured with percent accuracy by current
experiments: the solar $\Delta m_{21}^2 \equiv m^2_2 - m^2_1 \simeq
7.6\times 10^{-5}\,\eV^2$ and the atmospheric $|\Delta m_{31}^2| \equiv
|m^2_3 - m^2_1|\simeq 2.5\times 10^{-3}\eV^2$ mass gaps. Matter
effects in the sun tell us that the mass eigenstate with the larger
electron neutrino fraction has the smaller mass. We identify this state with ``1'' and
the heavier state (with a smaller electron neutrino fraction) with
``2''. Therefore, the solar mass splitting is positive. However, current neutrino oscillation data are unable to determine the sign of the
largest mass splitting, the atmospheric mass gap. Two possible scenarios therefore appear, 
corresponding to the two possible signs of $\Delta m_{31}^2$:
the normal hierarchy (NH hereafter), in which the atmospheric gap
is positive, and corresponds to $m_1<m_2<m_3$, and the inverted
hierarchy (IH in what follows), in which the atmospheric gap is negative, and corresponds to $m_3<m_1<m_2$~\footnote{Recent results for the NOvA long
  baseline experiment show that the best fit is obtained for the NH
  scheme, and rule out at $\sim 3\sigma$ a small region of the IH
  scenario, for some particular ranges of the mixing parameters. However,
a large fraction of the IH region is still allowed. Antineutrino data
can shed light on these results, see \url{http://nusoft.fnal.gov/nova/results/index.html}~\cite{nova}.}.
Assuming that the mass of the lightest mass eigenstate is
 zero, which equals to set to zero the mass of $m_1$ ($m_3$) in the NH (IH), it is
 possible to obtain a lower bound on the sum of neutrino masses of
 $M_\nu =\sqrt{\Delta m_{21}^2}+\sqrt{\Delta m_{31}^2}\simeq0.06\,\eV$ ($M_\nu=\sqrt{\Delta m_{31}^2}+\sqrt{\Delta m_{31}^2+\Delta m_{21}^2}\simeq0.1\,\eV$) from neutrino oscillation measurements. 
 
Neutrino mixing phenomena are sensitive to the neutrino mass
 splittings only, not to the individual neutrino masses nor to the
 overall mass scale.
Cosmology provides one of the most suitable places where to test and extract
the neutrino mass ordering~\cite{Lesgourgues:2004ps,Slosar:2006xb,Jimenez:2010ev,Hamann:2012fe,Hall2012MNRAS}, see also the recent work of
Refs.~\cite{Hannestad:2016fog,Xu:2016ddc}. Despite the fact that current bounds on
the neutrino mass $M_\nu$ show a dependence on how the mass is
distributed among the three mass eigenstates~\cite{Giusarma:2016phn}, 
present cosmological measurements are not able to firmly single out
nature's choice for the mass hierarchy. Consequently, in the
absence of a robust measurement of the neutrino mass ordering, a
desirable bound on $M_\nu$ would be one which does not rely on any assumption (or, to be more precise: that relies on the less informative possible assumption) 
about the hierarchical distribution of the total mass among the three eigenstates. 
This kind of problem, where the distribution of the parameters of the model under scrutiny are themselves conditionally dependent on the 
so-called \textit{hyperparameters} (namely, the bounds on $M_\nu$ are extracted by assuming a specific mass splitting), is a typical example of a hierarchical model in statistical inference. 
In this work, we propose
a novel method to get a hierarchy-independent bound on $M_\nu$, by means of a new discrete parameter, the \textrm{hyperparameter},
$\htype$ (that can in practice be identified with the sign of the atmospheric mass splitting), introduced in the standard Monte Carlo Markov chain (MCMC) analysis. This innovative strategy benefits from the fact
that the sensitivity to the neutrino hierarchy is simply and unbiasedly extracted from
the posterior probability distribution of $\htype$. We shall add this parameter while analyzing
current and future cosmological data, to illustrate the power of this technique. 

We stress that our approach is different from the one, already found in the literature, in which a single, continuous parameter
is used to parametrize both the sum of neutrino masses and the hierarchy \cite{Jimenez:2010ev}. The drawback
of the approach proposed in Ref. \cite{Jimenez:2010ev}, when used in an MCMC framework\footnote{While this paper was being finalized, Ref. \cite{Xu:2016ddc} appeared, using the parameterization proposed in Ref. \cite{Jimenez:2010ev} in an MCMC framework to derive constraints on the neutrino hierarchy.}, is that it is not possible to disentangle the
prior assumptions on the individual masses and on the hierarchy; in particular, a flat (uninformative) prior on the hierarchy
implies a non-flat prior on the mass. In our approach, we are free to specify noninformative priors for both the hierarchy and the
mass of the lightest eigenstate.

Apart from cosmological probes, there also exist laboratory avenues which are sensitive to the absolute mass scale. In this context, neutrinoless double $\beta$ decay ($0\nu2\beta$) searches (see e.g. Refs.~\cite{Avignone:2007fu,Cremonesi:2013vla,GomezCadenas:2011it,Dell'Oro:2016dbc}) are intriguing, as a positive signal would guarantee that neutrinos have a non-zero Majorana mass \cite{Schechter:1981bd}. Double
beta decay is a rare spontaneous nuclear transition in which the
charge of two isobaric nuclei changes by two units, emitting two
electrons. The dominant mode of this decay also produces two electron
antineutrinos, conserving lepton number and therefore, it is
allowed in the standard model framework. Double $\beta$ decay without
antineutrino emission, violating lepton number by two units, is the neutrinoless double $\beta$ decay. Planned $0\nu2\beta$ experiments might have the required sensitivity to completely cover the region of the parameter space where a positive signal is expected in the case of IH distribution of the total neutrino mass. Robust limits on the total neutrino mass coming from cosmology can further reduce the allowed region of the parameter space where to look for $0\nu2\beta$ events.

The paper is organized as follows: we describe our method and provide
details of the parameterization we adopt in Sec.~\ref{sec:method}; 
we present and discuss the implications for present cosmological data,
future CMB and BAO missions and neutrinoless double beta decay
experiments in Secs.~\ref{sec:current}, \ref{sec:4cast} and
\ref{sec:0n2b} respectively. We conclude in Sec.~\ref{sec:conclusion}.

\section{Method and data}\label{sec:method}
\subsection{Statistical framework and choice of the relevant parameters}

The problem we deal with in this work is a typical example of a statistically hierarchical\footnote{The meaning of ``hierarchical'' here has not to be confused with the two different neutrino mass distributions, or hierarchies. While we shall make use of the same terminology to refer to different concepts throughout the text, the context in which it is employed will help solving the ambiguity.} model (see e.g. \cite{bayes}). 
A key feature of hierarchical problems is that the parameters $\vec{\theta}$ of the model introduced for constraining the 
observables through the data $\vec{d}$ are modeled conditionally on further parameters, the \textit{hyperparameters} $\vec{\phi}$, which have themselves their own prior probability distribution $p(\vec{\phi})$. As a result, 
we can define a joint prior distribution
\begin{equation}
\Pi \equiv p(\vec{\phi},\vec{\theta})=p(\vec{\phi})p(\vec{\theta} | \vec{\phi})
\end{equation}
so that the proper posterior distribution $\mathcal{P}\equiv p(\vec{\theta},\vec{\phi}\,|\,\vec{d})$ of the parameters (both ``normal'' and ``hyper'')
can be written, using Bayes' theorem, as
\begin{equation}
\mathcal{P} =  \frac{\Pi \cdot \mathcal{L}}{\mathcal{E}} \propto p(\vec{\phi},\vec{\theta})p(\vec{d}|\vec{\phi},\vec{\theta})=p(\vec{\phi},\vec{\theta})p(\vec{d}|\vec{\theta})
\end{equation}
where $\mathcal{L} \equiv p(\vec{d}\,|\,\vec{\theta},\vec{\phi})$ is the likelihood function, in which we dropped the explicit dependence on $\vec{\phi}$, since the data depends on $\vec{\phi}$ only through $\vec{\theta}$, and $\mathcal{E} \equiv \int \mathcal{L}\cdot\Pi \,d\vec{\theta}d\vec{\phi}$ is the model evidence, or marginal likelihood. The latter does not depend on the parameters, and thus represents just a multiplicative constant as long as parameter estimation is concerned\footnote{We note that, even though
in this paper we are in principle also addressing a problem of model selection - i.e., determining the correct model for neutrino hierarchy -, the use of
the hyperparameter allows to map this into a parameter estimation problem.}.

In the case under investigation in this work, the model parameters are extracted conditionally on the choice of the 
neutrino mass hierarchy. This choice is modelled by introducing a discrete hyperparameter $\htype$ that can take two values, corresponding to NH and IH (i.e to $\sgn \left(\Delta m^2_{31}\right) = + 1$ or $-1$, respectively). Since little is known from current experiments about the preference for one of the two neutrino hierarchies, either normal or inverted, we assign equal \textit{a priori} probability to the two possible outcomes that $\htype$ could take.

We therefore perform a MCMC analysis of an eight-dimensional
parameter space. We consider the usual set of six cosmological
parameters in the $\lcdm$ scenario, namely the baryon density $\obh$,
the cold dark matter density $\och$, the angular size of the sound
horizon $\theta_s$, the reionization optical depth $\tau$, the scalar
spectral index $n_S$ and the amplitude $\loga$ of the power spectrum
of primordial scalar perturbations normalized at the pivot scale
$k_0=0.05\,\mathrm{Mpc^{-1}}$. All these parameters are extracted from flat prior distributions.

The inclusion of massive neutrinos
is perfomed in the following way: we assume three massive
non-degenerate eigenstates sharing the same temperature $T_\nu=(4/11)^{1/3}T_\gamma$. We sample,
again with a flat prior\footnote{Since the relation between $m_\mathrm{light}$ and $M_\nu$ is nonlinear, this
is in principle different than sampling over $M_\nu$, as it is usually done.},
over the lightest eigenstate mass $m_\mathrm{light}$ (thus corresponding to the seventh parameter of the model), which is
equivalent to $m_1$ ($m_3$) in the NH (IH) scenario. The mass of the
remaining two neutrino states is set by oscillation measurements
through the solar and atmospheric mass gaps, i.e. the so-called
squared mass differences, defined as $\Delta m_{ij}^2=m_i^2-m_j^2$,
with $i, j=1, 2, 3$. While the solar mass splitting is by convention
positive, i.e. $\Delta m_{21}^2>0$, the sign of the
atmospheric mass gap $\Delta m_{31}^2$, as previously stated, remains
still unknown, and it depends on the hierarchical distribution of the total mass
among the eigenstates, with $\Delta m_{31}^2>0(<0)$ in the NH (IH)
scenario. This is the reason why the lightest eigenstate corresponds to $m_1$
in the NH scenario, while it is mapped onto $m_3$ in the IH scenario. We use the latest best-fit values for
the oscillation mass gaps ~\cite{nufit,Forero:2014bxa}. 

As anticipated at the beginning of this section, the choice of the hierarchy is addressed
via the discrete \textrm{hyperparameter} $\htype$, which is the eight parameter of the model. 
At each step, we extract $\htype$ from $\{\mathrm{NH},\,\mathrm{IH}\}$, 
assigning equal \textit{a priori}
probability to the two hierarchical scenarios (i.e., we use the discrete equivalent of a flat prior). 
In the formalism sketched at the beginning of the section, $\vec{\theta} \equiv (\obh,\,\och,\,\theta_s,\,\tau,\,n_s,\,\loga,\,m_\mathrm{light})$,
while $\vec\phi = \htype$.
The inclusion of the 
\textrm{hyperparameter} allows us to handle our ignorance about 
the true hierarchical distribution of the mass as a nuisance
parameter, to be marginalized over. 
In this way, the posterior distribution of $\htype$ for a given datasets contains 
information about the preference for one of the two hierarchies from that dataset. This is easily done from the chains generated by the MCMC algorithm, 
by computing the marginalized probabilities $\mathcal{P}_\mathrm{NH}$ and $\mathcal{P}_\mathrm{IH}$, defined as
\begin{equation}\label{eq:pnh}
\mathcal{P}_\mathrm{NH} \equiv p(\htype=\mathrm{NH}\, |\, \vec{d}) = \int \mathcal{P}\left(\vec\theta,\,\htype = \mathrm{NH}\right) d\vec\theta \, ,
\end{equation}
and similarly for $\mathcal{P}_\mathrm{IH}$. This information is conveyed by reporting the ``odds'' for NH vs. IH,
i.e. the ratio $\mathcal{P}_\mathrm{NH}~:~\mathcal{P}_\mathrm{IH}$\footnote{We would like to report that Eq.~\ref{eq:pnh} is equivalent to Eq. 2.1 of~\cite{Hannestad:2016fog}, as it should be from the application of the basic rules of probability (including Bayes’ theorem). The novelty of our approach is that we include the parameter describing the hierarchy directly in the Monte Carlo, together with $m_\mathrm{light}$. This means that we don’t need to assume that the likelihood only depends on $M_\nu$. This is certainly a well-motivated approximation for present data. However our method can also be applied in cases in which this approximation does not work anymore (like for future experiments, or when non-cosmological data are added to the analysis). Another advantage is that, in our approach, we can include a flat prior on $m_\mathrm{light}$. We also obtain, for free, limits on the other parameters that take into account the uncertainty on the hierarchy. So, even though the starting point of the two approaches (Eq.~\ref{eq:pnh} of this work and Eq. 2.1 of ~\cite{Hannestad:2016fog}) is the same, the implementation is different, and, in our case, more general. Nevertheless, we would like to emphasize that, when the two approaches are expected to lead to the same results - as it is the case for present data that are only sensitive to the sum of the masses -, they actually do.}.

To compute the cosmological constraints and the 
posterior probability distributions in this
extended $\lcdm$ scenario,  we make use of the latest version of the publicly available MCMC package \texttt{cosmomc}
~\cite{Lewis:2002ah,Lewis:2013hha}, exploiting the Gelman
and Rubin statistics for monitoring the convergence of the generated chains~\cite{An98stephenbrooks}. We quote
our results in terms of 95\% credible intervals for the parameters. Given that we will be dealing
with possibly multimodal distributions, the credible intervals can consist of the union of disjointed regions.

\subsection{Cosmological Datasets}

Current CMB and BAO measurements are considered. Results are presented
separately for CMB temperature and low-multipole polarization data
(the former ranging from multipoles $\ell=2$ up to $\ell=29$) from the
Planck mission (TT + lowP), and for the addition, to the previous
measurements, of the high multipole (i.e. small-scale) polarization and
cross-correlation spectra, i.e. the full Planck data release
(TT, TE, EE + lowP), see Refs.~\cite{Ade:2015xua,Adam:2015rua,Aghanim:2015xee}. We remind the reader
that since, as discussed in Refs.~\cite{Ade:2015xua,Aghanim:2015xee}, small-scale polarization 
data could still be affected by low-level residual systematics,
results obtained without using them should be regarded as more reliable.
These measurements are analyzed by
means of the publicly available Planck likelihood code~\cite{Aghanim:2015xee}, and
foregrounds or extra nuisance parameters are carefully treated
following the prescription detailed in Refs.~\cite{Ade:2015xua,Aghanim:2015xee}.

We combine the two CMB datasets described above with geometrical information from 
galaxy clustering, i.e. via the BAO
signature. All the BAO measurements we exploit here are expressed as determinations of
$D_{\textrm{V}}(z_{\textrm{eff}})/r_{\textrm{s}}(z_{\textrm{drag}})$,
with 
\begin{equation}
D_{\textrm{V}}(z)= \left[(1+z)^2 D_A(z)^2\frac{z}{H(z)}\right]^{1/3}
\end{equation}
representing a combination of the line-of-sight clustering information
(as encoded by the Hubble parameter $H$) and the transverse clustering
information (encoded in the angular diameter distance $D_A$)
at the effective redshift $z_\mathrm{eff}$ of the survey, and
$r_{\textrm{s}}(z_{\textrm{drag}})$ being the sound horizon at the drag
epoch\footnote{The drag epoch is defined as the time at which baryons are released from 
the Compton drag of the photons, see Ref.~\cite{Eisenstein:1997ik}.}. Concretely, 
we make use of the BAO results from the 6dF Galaxy Survey
(6dFGS)~\cite{Beutler:2011hx} and from the BOSS DR11 LOWZ and CMASS samples~\cite{Anderson:2013zyy}, focusing at $z_{\textrm{eff}}=0.106$,
and $z_{\textrm{eff}}=0.32, 0.57$,
respectively.  The combination of these measurements will be referred
to as BAO. The addition of galaxy clustering measurements, apart from
breaking pure \textit{geometrical} degeneracies among different cosmological parameters
(as, for instance, the one existing between the Hubble constant 
$H_0$ and the neutrino mass $M_\nu$ ~\cite{Giusarma:2012ph}), also helps
enormously in pining down the neutrino mass limits, as the free
streaming nature of sub-eV neutrinos will leave a clear imprint in the
galaxy power spectrum at scales in the regime of interest, see e.g.~\cite{Giusarma:2016phn}.

We also perform forecasts for future CMB and galaxy
clustering data. For the fiducial values of the six $\Lambda$CDM parameters, 
we use the mean values of the estimates reported in \cite{Aghanim:2016yuo} for 
PlanckTT+SIMlow. Concerning CMB measurements, we consider a
future COrE-like~\cite{Bouchet:2011ck} satellite mission, generating mock lensed temperature
and polarization data accordingly to Refs.~\cite{Bond:1998qg,Bond:1997wr}. We assume
perfect foreground subtraction as well as precise control of systematics. The expected noise
spectra (which relies on specific experimental setup, such as
the sky fraction, the beam width, and the temperature and polarization
sensitivities) has been verified against previous results in the
literature~\cite{Errard:2015cxa,Escudero:2015wba}, finding an excellent agreement. 
 Future galaxy clustering data are added by means of the expected 
 independent observations of the BAO signal along and across the line
 of sight from the Dark Energy Instrument (DESI) Experiment~\cite{Levi:2013gra}.
DESI observations will provide separate measurements of $H(z) r_{\textrm{s}}$
and $D_A(z)/r_{\textrm{s}}$ at a number of redshifts. This survey is
expected to cover $14000$ squared degrees of the sky in the
redshift range $0.15 < z < 1.85$. We follow the DESI Conceptual Design
Report, and also Ref.~\cite{Font-Ribera:2013rwa} for the percentual errors on both $H(z) r_{\textrm{s}}(z)$
and $D_A(z)/r_{\textrm{s}}(z)$ expected from the three types of DESI tracers,
 (namely, Emission Line Galaxies, Luminous Red Galaxies and High
 Redshift Quasars) and assume an identical 0.4 correlation coefficient between the percentual errors on $H(z) r_{\textrm{s}}(z)$
and $D_A(z)/r_{\textrm{s}}(z)$ (see e.g. \cite{Seo:2007ns,Font-Ribera:2013rwa}). 

Furthermore, we include, in the future cosmological data, a $1\%$-measurement of the 
Hubble constant $H_0$. This is implemented in the form of a gaussian prior on $H_0$
in the analysis with future cosmological data. However, given the
fact that the COrE sensitivity is such that it will expectedly allow to determine the value of $H_0$ below that precision, 
such a prior on the Hubble constant does not play any crucial role.

\section{Results from present cosmological measurements}
\label{sec:current}
\begin{table*}
\begin{center}\footnotesize
\scalebox{1.04}{\begin{tabular}{lccccc}
\hline \hline
         & TT + low P &         TT, TE, EE + low P &         TT + lowP + BAO &         TT, TE, EE + low P +  BAO\\
\hline
\hspace{1mm}\\
$ M_\nu$ [eV] &$[0.058 - 0.740]$ & $[0.058 - 0.558]$ & $[0.058 - 0.232]$ & $[0.058 - 0.200]$ \\
\hspace{1mm}\\
$m_\textrm{light}$ [eV] & $<0.244$ & $<0.183$& $<0.0695$ &$<0.0577$  \\
\hspace{1mm}\\
$m_1$  [eV] &$<0.246$ & $<0.186$ & $<0.079$ & $<0.068$ \\
\hspace{1mm}\\
$m_2$  [eV] &$[0.009 - 0.246]$  &$[0.009 - 0.186]$  & $[0.009 - 0.079]$  &  $[0.009 - 0.069]$\\
\hspace{1mm}\\
$m_3$  [eV] & $ < 0.243 $&$ < 0.185$& $ < 0.082$&  $ < 0.072$ \\
\hspace{1mm}\\
$\Omega_c h^2$  & $0.1205^{+0.0046}_{-0.0045}$ &$0.1203\pm0.0030$&$0.1184^{+0.0026}_{-0.0027}$&$0.1189\pm0.0022$\\
\hspace{1mm}\\
$H_0\,\mathrm{[km\,s^{-1}\,Mpc^{-1}]}$  &$65.1^{+4.0}_{-5.3}$&$65.6^{+2.8}_{-3.7}$&$67.4\pm1.2$&$67.2\pm1.1$\\
\hspace{1mm}\\
$\Omega_m$ & $0.346^{+0.081}_{-0.059}$&$0.337^{+0.052}_{-0.039}$ & $0.313^{+0.016}_{-0.015}$ & $0.316\pm0.014$\\
\hspace{1mm}\\
\hline
\hspace{1mm}\\
$\htype$ odds (NH:IH) & $1:1$ &$9:8$ &$4:3$ & $3:2$\\
\hspace{1mm}\\
\hline
\hline
\end{tabular}}
\caption{$95\%$~credible intervals for the total neutrino mass, the mass of the lightest neutrino eigenstate and the individual neutrino masses, as well as for other cosmological parameters, for different combinations of current CMB and BAO data. All the bounds reported here also take into account information from oscillation measurements. In the last row, we quote the odds for the NH vs. the IH scenario.}
\label{tab:current}
\end{center}
\end{table*}

In this section, we present the bounds on the neutrino mass parameters derived from different combinations
of current cosmological probes, as well as discuss the sensitivity of the very same probes to the neutrino mass ordering.
 Table~\ref{tab:current} shows the $95\%$~CL constraints on the total
 neutrino mass, on the lightest neutrino mass, and on the individual
 neutrino masses associated to each neutrino mass eigenstate after
 marginalization over the $\htype$ parameter. For each parameter that appears in the table, with the exception of $\htype$,
we quote our results in the following way: if i) the 95\% confidence interval includes one of the edges of the prior range for that parameter (this is the case for $M_\nu$ and
for the individual masses), 
or ii) the posterior  probability distribution is bimodal (this is sometimes the case for $M_\nu$, see below), 
then we report the 95\% confidence interval in the form $[\mathrm{min},\,\mathrm{max}]$;
iii) otherwise, we report the 95\% confidence interval in the form $(\mathrm{mean}\pm\mathrm{uncertainty})$.
In the last line of the table, we report the results for $\htype$, in the form (odds for NH : odds IH).
The odds shown in the second column of Table~\ref{tab:current} show how CMB temperature data alone are not sensitive to the different mass
parameterizations. This is due to the broad bound that the CMB
temperature data set on the total neutrino mass, $M_\nu<0.740$~eV at $95\%$~CL.
In fact, the potential for cosmological observations to discriminate between
the two mass orderings mainly relies on the capability to push the upper limit on $M_\nu$ close or even below $0.1$~eV,
the minimal value of the mass allowed by oscillation data in the case of IH. 
This is mainly due to the fact that the region $M_\nu < 0.1\,\eV$ is only allowed in the case of normal hierarchy, so
that tighter upper bounds on $M_\nu$ end up favouring the NH scenario simply because of the larger volume in parameter space available
to the model. Moreover, the region of masses with $M_\nu \simeq 0.1\,\eV$ is the one where the mass patterns predicted by NH and IH,
and the resulting cosmological perturbations, differ the most. The differences are however small, given the sensitivity
of present, and possibly also future, experiments, and the dominant contribution to the constraining power still comes
from the sheer amount of volume in parameter space available to the two models.
On the opposite, when $M_\nu \gg 0.1\,\eV$ (i.e., $M_\nu \gg \Delta m^2_{21},\, \Delta m^2_{31}$), 
we are in a situation in which both hierarchies effectively coincide with the degenerate scenario 
$m_1\simeq m_2\simeq m_3$, and the differences in the evolution of perturbations induced by the mass ordering
are too small to have any observable consequence, given current sensitivities. 
Given that the we find a bound from CMB temperature anisotropies and large-scale polarization $M_\nu<0.740$~eV,
most of the parameter space available, given this data, is in the ``effectively degenerate'' region, where the two hierarchies 
cannot be distinguished .
Indeed, the posterior distributions for the mass fractions $f_{\nu,i} \equiv m_i/M_\nu$ for the
case of CMB temperature data are clearly
peaking on $f_{\nu,i}=1/3$, as it would be in the fully degenerate
scenario. Notice that the addition of small-scale CMB polarization
measurements slightly improves the neutrino mass bound ($M_\nu<0.558$~eV at
$95\%$~CL) but it does not change significantly the overall picture. 
 
We note that the bounds we find seems to be larger, when a direct comparison is possible, than those found in Ref.~\cite{Ade:2015xua} without
the marginalization over the two possible mass orderings: compare, e.g., our bound $M_\nu<0.740$~eV with $M_\nu<0.715\,\eV$, the 95\% upper bound from Planck TT+lowP in the context of the $\lcdm+M_\nu$ model, assuming three massive degenerate neutrinos~\cite{Ade:2015xua}. The reason is the following. In the present analysis, a non-vanishing lower bound on the total neutrino mass, $M_\mathrm{\nu,min}=0.058\,\eV$ is naturally imposed by taking into account neutrino oscillation measurements, 
while in Ref.~\cite{Ade:2015xua} it is only assumed that $M_\nu \ge 0$. As a consequence, in our case the 95\% confidence regions for $M_\nu$ are shifted,
by definition, towards larger masses. The same care should be applied when comparing to similar constraints reported in the literature.

\begin{figure*}
\begin{center}
\includegraphics[width=1.0\textwidth]{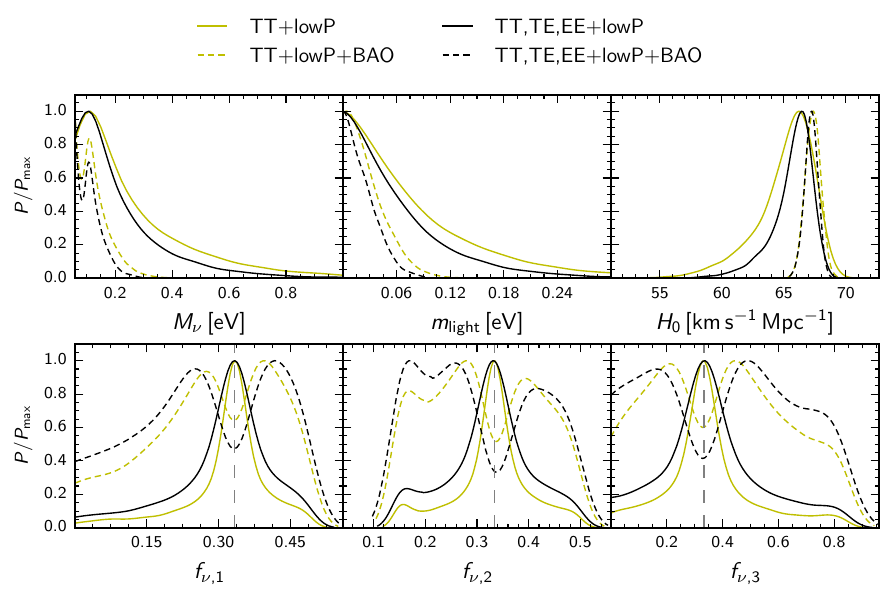}
\end{center}\caption{One-dimensional probability posterior distribution of a selection of parameters analyzed in this work, for the combinations of current CMB and BAO datasets reported in the top legend. In the top panels, we report the posterior distributions for the sum of the neutrino masses $M_\nu=\Sigma_i m_i$, where the index $i=1,2,3$ runs over the three mass eigenstates $m_i$; the mass carried by the lightest eigenstate $\mlight$ and the Hubble constant $H_0$. In the bottom panels, we report the posterior distributions for the neutrino mass fractions $f_{\nu,i}=m_i/M_\nu$. The solid (dashed) lines are for CMB alone (CMB plus BAO) measurements. The vertical dashed lines in the bottom panels refer to the expected value of $f_{\nu,i}=1/3$ in the case of a fully degenerate mass spectrum.  All the posterior shown in this figure also take into account information from oscillation measurements.}\label{fig:current}
\end{figure*}

The inclusion of BAO measurements results in much tighter neutrino mass
bounds than those
obtained with CMB data only. The combination of the cosmological data
then starts to be sensitive to the region of $\mlight$ in which the NH and the IH
scenarios correspond to different neutrino mass spectra. This can
be clearly noticed from the bimodal distributions in both $M_\nu$ and
in the neutrino mass fractions depicted in Fig.~\ref{fig:current}, 
where two distinct peaks appear, each one being associated to the most probable
value of the parameter \textit{for a given choice of the hierarchy}. 
In fact, focusing on $M_\nu$, we find that the most probable value is 
$M_\nu=0.059\,\eV=M_{\nu,\mathrm{min}}^\mathrm{NH}$, but a second peak is also clearly visible in $M_\nu=0.098\,\eV=M_{\nu,\mathrm{min}}^\mathrm{IH}$.
These peaks are associated to the (single) peak at $m_\mathrm{light} = 0$ in the posterior for the mass of the lightest neutrino,
that gets mapped to two distinct values of $M_\nu$ depending on the hierarchy. 

By focusing on the probability odds of the
\textit{hyperparameter} $\htype$ in the two hierarchies, one might be able to assess whether current cosmological data favour one of the two hierarchical scenarios and to what extent. We can conclude that while there is still no compelling
evidence for the cosmological data to prefer one of the two scenarios, 
the combination of CMB and BAO slightly favors the NH scheme (4:3 odds in favor of NH, without
using small-scale polarization, or 3:2 if we use it). This is also confirmed by the inspection of e.g. the top upper panel of Fig.~\ref{fig:current}, where the combination of CMB and BAO data is able to unveil a bimodal posterior distribution of $M_\nu$.

Finally, we note that the mean values and errors of the standard $\lcdm$ parameters
$\Omega_c h^2$, $H_0$ and $\Omega_m$ shown in Tab.~\ref{tab:current} are indeed very close to
those quoted in Ref.~\cite{Ade:2015xua} for the corresponding data
sets, and derived without taking into account the uncertainty on the hierarchy\footnote{The cosmological parameters constraints for the $\Lambda$CDM model and many extensions, from several combinations of Planck 2015
and external data, can be downloaded from the Planck Legacy Archive \url{http://www.cosmos.esa.int/web/planck/pla}.}. This is yet another 
reflection of the fact that in the most part of the high-probability region of the parameter space, the mass spectrum is effectively degenerate.

To conclude this section, current cosmological data are only mildly sensitive to the neutrino mass ordering, with a preference of 4:3 in favour of NH from the combination of Planck CMB data and BAO measurements.

\section{Forecasts for future CMB and BAO surveys}\label{sec:4cast}
In this section, we present forecasted constraints from future cosmological surveys on the neutrino mass parameters and discuss whether the improved sensitivity of the next-generation cosmological observatories will help unraveling the dilemma about the neutrino mass ordering. 

\begin{table*}
\begin{center}\footnotesize
\scalebox{1.04}{\begin{tabular}{lcccc}
\hline \hline
         & COrE: $M_\nu$= 0.1 eV, NH&         COrE: $M_\nu$= 0.1 eV,
         IH&  COrE: $M_\nu$= 0.06 eV\\
         & (+ DESI)& (+ DESI)& (+ DESI)\\
\hline
\hspace{1mm}\\
$ M_\nu$ [eV] &$[0.058 - 0.188]$  & $[0.058 - 0.186]$ &$[0.058 - 0.155]$\\
     &$(0.112^{+0.037}_{-0.040})$  & $(0.113^{+0.038}_{-0.042})$ &$([0.058 - 0.109])$\\
\hspace{1mm}\\
$m_\textrm{light}$ [eV] &$<0.0529$&$<0.0523$&$<0.0405$\\
&$(<0.0362)$&$(<0.0366)$&$(<0.0225)$\\
\hspace{1mm}\\
$m_1$  [eV] &$<0.067$  & $<0.068$ &$<0.0571$\\
&$([0.002 - 0.061])$  & $([0.002 - 0.061])$ &$(<0.0491)$\\
\hspace{1mm}\\
$m_2$  [eV] &  $[0.009 - 0.0664]$&$[0.009 - 0.0659]$&$[0.009 - 0.0577]$\\
&$([0.009 - 0.0555])$&$([0.009 - 0.0562])$&$([0.009 - 0.0499])$\\
\hspace{1mm}\\
$m_3$  [eV]  &  $ < 0.070$&$<0.070$&$< 0.063$\\
&$(< 0.061)$&$(<0.062)$&$(< 0.055)$\\
\hspace{1mm}\\
$\Omega_c h^2$  
&$0.1209^{+0.0011}_{-0.0010}$&$0.1209^{+0.0010}_{-0.00098}$&$0.12117\pm0.00087$\\
&$(0.12072\pm0.00058)$&$(0.12071^{+0.00058}_{-0.00057})$&$(0.12079^{+0.00054}_{-0.00052})$\\
\hspace{1mm}\\
$H_0\,\mathrm{[km\,s^{-1}\,Mpc^{-1}]}$& $66.27^{+0.95}_{-0.99}$&$66.30^{+0.93}_{-0.97}$&$[65.59 - 67.08]$\\
 & $(66.46^{+0.52}_{-0.48})$ &$(66.46^{+0.53}_{-0.50})$ &$(66.72^{+0.35}_{-0.44})$\\
\hspace{1mm}\\
$\Omega_m$ & $0.329^{+0.014}_{-0.013}$&$0.328^{+0.014}_{-0.013}$ &
$[0.318 - 0.339]$ &\\
 & $(0.3262^{+0.0066}_{-0.0070})$&$(0.3262\pm0.0070)$ & $(0.3230^{+0.0061}_{-0.0049})$ &\\
\hspace{1mm}\\
\hline
\hspace{1mm}\\
$\htype$ odds & $1:1$ &$1:1$ &$3:2$ \\
&$(1:1)$ &$(1:1)$ &$(9:1)$\\
\hspace{1mm}\\
\hline
\hline
\end{tabular}}
\caption{As Tab.~\ref{tab:current} but for future measurements from
  the COrE CMB mission and for the DESI galaxy survey, and three
  different fiducial models: NH with $M_\nu=0.1$~eV, IH with
  $M_\nu=0.1$~eV and NH with $M_\nu=0.06$~eV (second, third and fourth
  columns).} 
\label{tab:future}
\end{center}
\end{table*}

\begin{figure*}
\begin{center}
\includegraphics[width=1.0\textwidth]{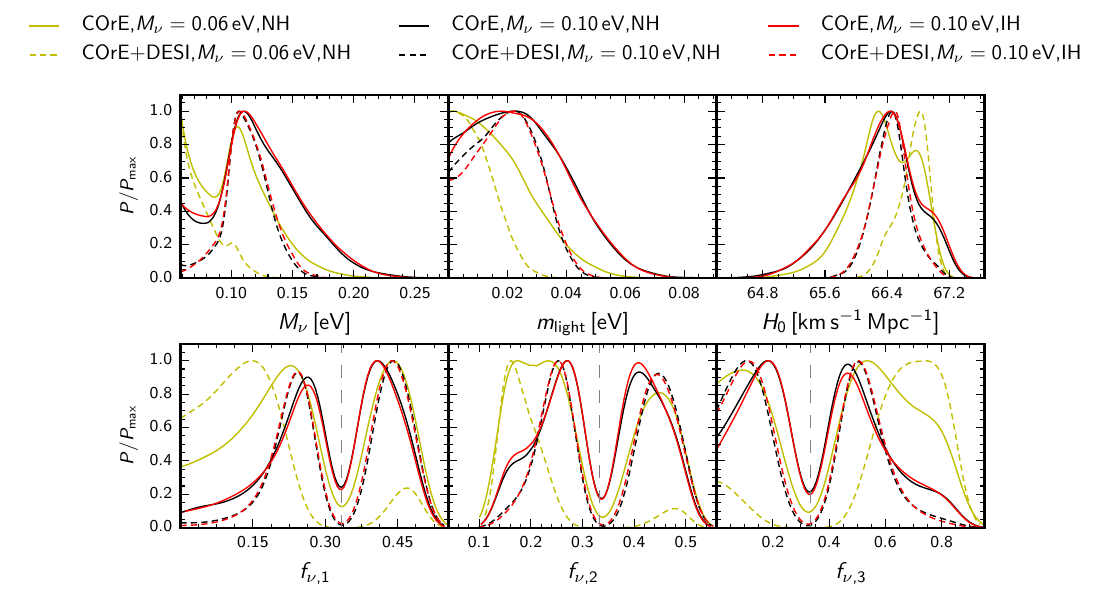}
\end{center}
\caption{One-dimensional probability posterior distribution of a
  selection of parameters analyzed in this work, for the combination
  of datasets reported in the figure. In the top panels, we report the posterior distributions for the sum of the neutrino masses $M_\nu=\Sigma_i m_i$, where the index $i=1,2,3$ runs over the three mass eigenstates $m_i$; the mass carried by the lightest eigenstate $\mlight$ and the Hubble constant $H_0$. In the bottom panels, we report the posterior distributions for the neutrino mass fractions $f_{\nu,i}=m_i/M_\nu$. The solid (dashed) lines refer
  to COrE (COrE plus DESI) forecasted MCMC results. The vertical dashed lines in the bottom panels refer to the expected value of $f_{\nu,i}=1/3$ in the case of a fully degenerate mass spectrum.  All the posterior shown in this figure also take into account information from oscillation measurements.}
\label{fig:futuref}
\end{figure*}

We present the results for the forecasted COrE-like~\cite{Bouchet:2011ck} CMB mission and a DESI-like survey~\cite{Levi:2013gra} in
Tab.~\ref{tab:future} and Fig.~\ref{fig:futuref}. The results are quoted following the same scheme adopted in Tab.~\ref{tab:current} and detailed at the beginning of Sec.~\ref{sec:current}. We have considered
three possible fiducial scenarios: two NH schemes, one with
$M_\nu=0.06\,\eV$ and the other one with $M_\nu=0.1\,\eV$, and one IH
scenario, also with $M_\nu=0.1\,\eV$. Notice that even for a future CMB mission as COrE
it will be very difficult to extract with high statistical significance the neutrino mass hierarchy in any
of the three fiducial scenarios explored here, even in the case of
$m_\textrm{light}=0$ ($M_\nu=0.06\,\eV$). Nevertheless, by a comparison between the
results in Tabs.~\ref{tab:current} and those in Tab.~\ref{tab:future}
one can learn that the expected sensitivity of COrE alone on the neutrino mass
measurements is slightly better than current combined CMB and galaxy
clustering searches. This accuracy could be reinforced, for instance,
by improved measurements from low-redshift experiments, such as adding a prior on the Hubble constant which simulates a $\sim
1\%$-measurement of $H_0$. However, the impact of such a constraint will be almost
negligible, as the COrE mission alone reaches already that precision in $H_0$.
We therefore focus on additional information coming from galaxy surveys and add future forecasted measurements of the Hubble
parameters and of the angular diameter distance from the DESI
survey~\cite{Levi:2013gra}. 

Adding BAO measurements improves
considerably the results for the fiducial model with
$M_\nu=0.06\,\eV$; in this case, we find a $9:1$ preference of NH versus IH. The
great improvement due to the addition of DESI BAO data when $M_\nu=0.06\,\eV$ can be clearly
visualized from the first two panels of Fig.~\ref{fig:futuref}: notice that
the second peak at $\sim 0.1\,\eV$ in the $M_\nu$ posterior is significantly reduced
after the inclusion of BAO data. 

On the contrary, for the $M_\nu=0.1\,\eV$ case, the situation is dramatically different.
In fact, even if the addition of BAO
measurements helps at pinpointing the value of sum of neutrino masses (as
it can be seen by comparing, in Fig.~\ref{fig:futuref}, the width of the dashed black and blue curves  
with that of their solid counterparts), nevertheless the data still remain completely uninformative
for what concerns the mass splitting, as it can be inferred by looking at the numbers reported in the last row of 
Tab.~\ref{tab:future}, second and third columns. This points to the fact that, as already explained in Sec. II in reference
to current experiments, the capability of future CMB and BAO observations to discriminate the neutrino mass hierarchy
mainly relies on volume effects, i.e., on the possibility of excluding $M_\nu \ge 0.1\,\eV$ with a high statistical significance;
this is the case for the fiducial model with $M_\nu \ge 0.06\,\eV$.
When instead $M_\nu =0.1\,\eV$ (or larger), as in the other two fiducial models considered here, the 
two mass orderings should be disentangled through the effect of the individual neutrino masses 
on the evolution of cosmological perturbations. Our findings clearly indicate that this is beyond the reach
of next-generation CMB and BAO experiments, even in the most optimistic case (for $M_\nu > 0.1\,\eV$,
the differences between the two hierarchies are even smaller). A possible improvement could come 
from highly accurate measurements
of the matter power spectrum \cite{Jimenez:2010ev}; see also Ref.~\cite{Pritchard:2008wy} for an appraisal of future
$21$~cm facilities.
Alternatively, one should combine results coming from cosmological analysis with constraints obtained in laboratory searches, as we shall see in the following section.

\section{Implications for $0\nu2\beta$}\label{sec:0n2b}
In the following, we shall discuss the implications of our analyses for
current and future neutrinoless double beta decay searches, as well as discuss the possibility 
to gain more information by the combination of cosmological observations with $0\nu2\beta$ experiments. 
 The non-observation of neutrinoless double $\beta$ decay processes 
provides at present bounds on the so-called \textit{effective Majorana mass}
of the electron neutrino
\begin{equation}
m_{\beta\beta} = \frac{m_e}{\mathcal{M}\sqrt{G_{0\nu}T^{0\nu}_{1/2}}}~,
\label{eq:thalf}
\end{equation}
where $T^{0\nu}_{1/2}$ is the neutrinoless double
$\beta$ decay half-life, $m_e$ is the electron mass, $G_{0\nu}$ is a
phase-space factor and $\mathcal{M}$ is the nuclear matrix element (NME),
a crucial quantity whose uncertainties affect significantly the interpretation of current and future
searches for $0\nu2\beta$ decay events.

The effective Majorana mass is related
to the neutrino mass eigenvalues as follows:
\begin{equation}
m_{\beta\beta} \equiv \left|\sum_{k} U^2_{ek} m_k\right| = \left|\sum_{k} e^{i\phi_k} V^2_{ek} m_k\right|~,
\label{eq:mbb}
\end{equation}
where we have written the mixing matrix $U$ of Eq. (\ref{eq:mix}) as the product of a matrix $V$, that contains
the mixing angles and the Dirac phase, and a diagonal matrix that contains the \textit{Majorana phases $\phi_k$}, see e.g. Sec. 14 of Ref.~\cite{pdg2016}. 
Since one of the phases can always be rotated away, we can assume that $\phi_1= 0$.
While the elements of $V$ are the same that enter the oscillation probabilities, 
the Majorana phases
play no role in neutrino oscillation processes.
However, they enter in the determination of the $0\nu2\beta$ half-life, as it is clear from Eqs. (\ref{eq:thalf})-(\ref{eq:mbb}),
and are thus crucial for $0\nu2\beta$  experiments. In fact, if the hierarchy is normal, the phases could also arrange to produce
a vanishing Majorana mass, and thus no observable $0\nu2\beta$ signal, even for non vanishing values of the individual masses.

In the previous sections, we have combined cosmological observations (that mainly constrain $M_\nu$) 
and oscillation measurements (that probe the mass differences) to derive limits on the individual masses,
taking into account our ignorance of the mass hierarchy. Since oscillation measurements also constrain 
the elements of the mixing matrix, we can exploit the same strategy to derive constraints on the Majorana mass 
[related to the other parameters of the MCMC analysis by Eq. (\ref{eq:mbb})], provided that we also
take into account our ignorance of the true values of the Majorana phases.

Following Ref.~\cite{Gerbino:2015ixa}, we thus consider the two Majorana phases as extra parameters in
the MCMC analysis, $\phi_{2}$ and $\phi_{3}$, with flat priors in the range $[0,\,2\pi]$,
as currently the values of these phases are totally unknown. We also do not consider here
the possibility of future independent measurements of the phases.
We then extend the analysis discussed in the previous sections for
Planck TT,TE,EE+lowP+BAO, as an example of current data, and for COrE+DESI, as an example of future data, 
extracting the posterior distribution for the Majorana mass.
In case of future data, we consider two fiducial models with NH mass ordering and either  $M_\nu=0.06\,\eV$ or $M_\nu=0.1\,\eV$.

\begin{figure}
\begin{center}
\includegraphics[width=0.9\columnwidth]{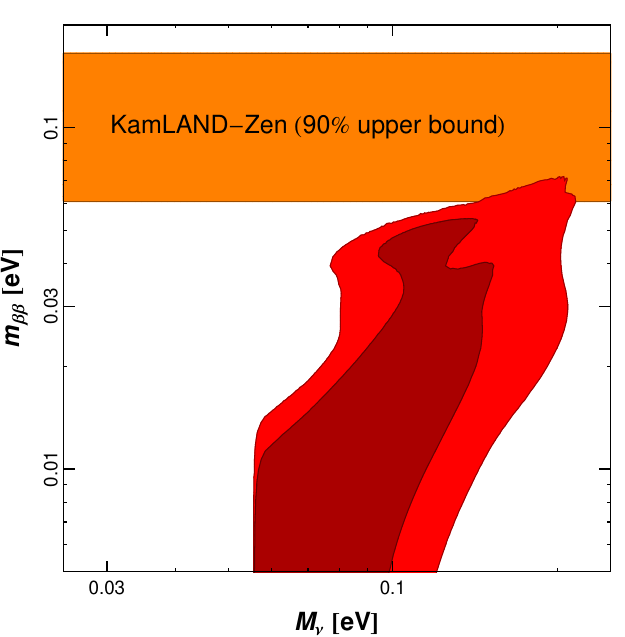}
\end{center}\caption{Two dimensional 68\% and 95\% probability contours in the $M_\nu-m_{\beta\beta}$ plane, for current cosmological
(Planck TT, TE, EE + BAO) and neutrino oscillation data. The contours are calculated by marginalizing over $\htype$, so they take into account the uncertainty on the mass ordering. The contours trace two distinct regions of large probability, that are more clearly visible in Fig.\ref{fig:densmbbmnu}, roughly corresponding to the portion of parameter space preferred by each of the two hierarchies (see the main text for details). The orange horizontal bands correspond to the 90\% upper bounds on $m_{\beta\beta}$
  obtained from KamLAND-Zen \cite{KamLAND-Zen:2016pfg}, for different assumptions for the values of the nuclear matrix elements that enter into the calculation of $m_{\beta\beta}$.\label{fig:betabeta}}
\end{figure}

\begin{figure}
\begin{center}
\includegraphics[width=0.9\columnwidth]{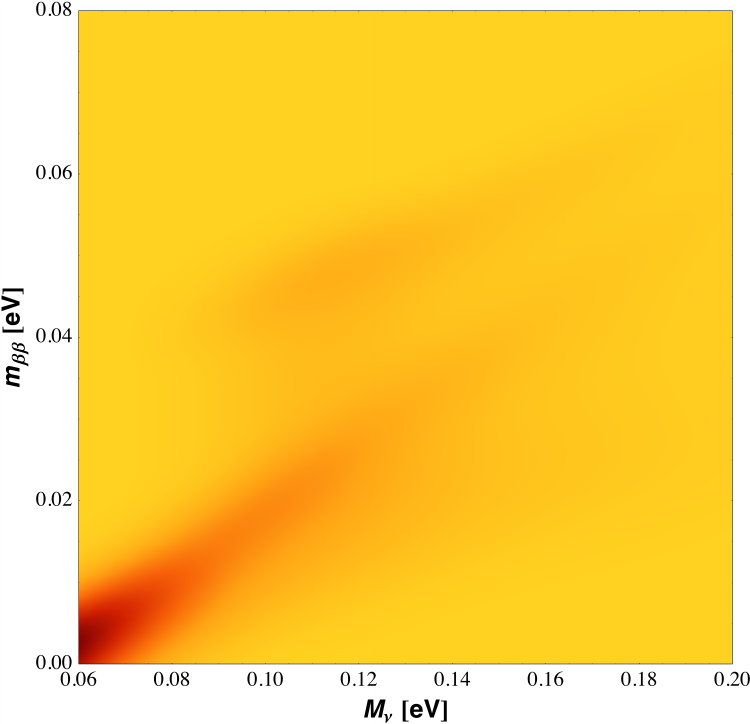}
\end{center}
\caption{Density plot of the two-dimensional posterior in the $M_\nu-m_{\beta\beta}$ plane, for current cosmological
(Planck TT, TE, EE + BAO) and neutrino oscillation data. Darker colors correspond to higher probability regions.
The plot shows that the posterior is bimodal. \label{fig:densmbbmnu}}
\end{figure}

 The $95\%$~CL limit we find from current cosmological data, after marginalization over the 
\textit{hyperparameter} $\htype$, is $m_{\beta\beta} <0.056\,\eV$. We illustrate in
Fig. ~\ref{fig:betabeta} the $68\%$ and $95\%$~CL probability contours in
the $M_\nu-m_{\beta\beta}$ plane. We also depict together the tightest bounds on
neutrinoless double beta decay searches, coming from the KamLAND-Zen
experiment, with $90\%$~CL limits on $m_{\beta\beta}<61-165$~meV\footnote{We remind the reader that the quoted confidence levels for the KamLAND-Zen experiment are drawn in the context of frequentist analysis. As a result, a bayesian analysis of KamLAND-Zen measurements should be conducted (see e.g.\cite{Gerbino:2015ixa,Caldwell:2006yj}) in order to perform properly a direct comparison of its constraining power and the combination of cosmological datasets discussed in this work. Thus, the (frequentist) limits from KamLAND-Zen shown in Fig.~\ref{fig:betabeta} have an illustrative purpose only.},
the precise value depending crucially on the NMEs assumption~\cite{KamLAND-Zen:2016pfg}. 
A visual inspection of the two-dimensional posterior makes evident that the posterior is bimodal,
as it can be seen more clearly in Fig. \ref{fig:densmbbmnu}.
These results show that there exist two separated regions of large probability, one preferring vanishingly small values of $m_{\beta\beta}$ and extending
down to $M_\nu = 0.06$~eV,
the other peaking around $m_{\beta\beta}= 0.04$~eV and $M_\nu =0.1$~eV. This is due to the fact that we are not assuming one of the two hierarchies,
but we are instead marginalizing over the mass ordering. The two regions of large probability roughly trace the portions of parameter space 
that would be preferred assuming either the normal or inverted hierarchy. To be more precise, the preference for $m_{\beta\beta}=0$ is given 
by models with normal hierarchy, while the region around $m_{\beta\beta}=0.04\,\eV$ is mostly due to models with inverted 
ordering (with some contribution from the tail of the posterior distribution of models with $\htype=\mathrm{NH}$). 

Interestingly, current sensitivities to $m_{\beta\beta}$
start to reach the allowed region by cosmological and oscillation measurements, and,
consequently, if nature has chosen the inverted hierarchy, a positive
signal from neutrinoless double beta decay searches could be
imminent (providing of course neutrinos possess a Majorana character
and barring highly exotic physical scenarios). At present, both
cosmological and laboratory tests of $m_{\beta\beta}$ provide very
similar constraints on the Majorana mass (assuming the most
favorable values for the NMEs).

We would like to emphasize again that the two large-probability regions in the $\{M_\nu,\,m_{\beta\beta}\}$ plane are not drawn separately by assuming, in turn, 
each of the two hierarchies as the true one, \textit{a priori}, as usually done in literature. On the contrary, the appearance of the two regions is a direct consequence of the hierarchical model built via the \textit{hyperparameter} $\htype$ and of the corresponding marginalization. This also allows to assess the relative probability of $m_{\beta\beta}$ lying in each of the two regions.

\begin{figure*}
\begin{center}
\includegraphics[width=0.5\textwidth]{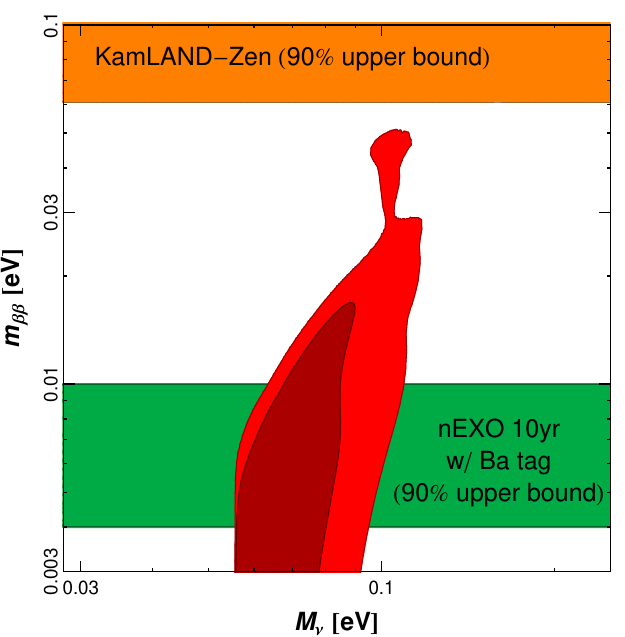}\includegraphics[width=0.5\textwidth]{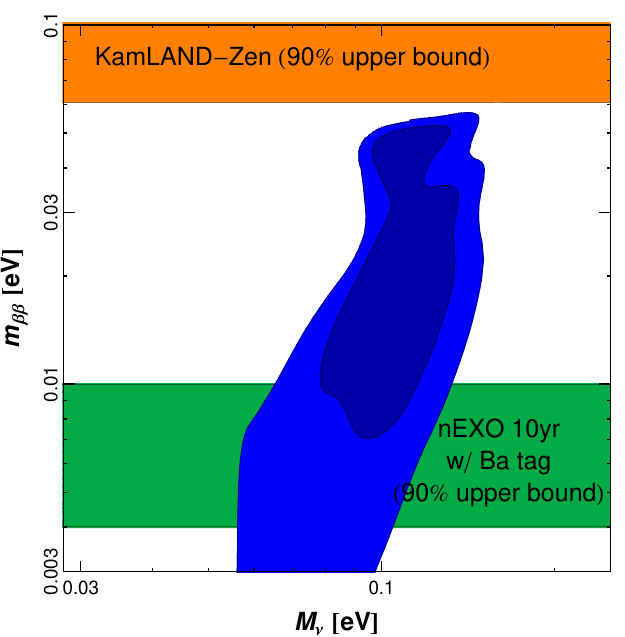}
\end{center}\caption{Two dimensional 68\% and 95\% probability contours in the
  $M_\nu-m_{\beta\beta}$ plane, for future cosmological (COrE+DESI) and neutrino oscillation data,
  considering the NH scenario as the nature's choice and $M_{\nu}=0.06\,\eV$ (left panel) or $M_{\nu}=0.1\,\eV$ (right panel). 
  The contours are calculated by taking into account the uncertainty on the mass ordering, without fixing a priori the mass hierarchy. The two regions that can be inferred in both figures (even if not completely isolated) correspond to the two mass orderings. The horizontal bands correspond to the 90\% upper bounds on $m_{\beta\beta}$
  obtained from KamLAND-Zen \cite{KamLAND-Zen:2016pfg} (orange) and to those expected from a future nEXO-like experiment (green) \cite{nEXO}
  (assuming a vanishing Majorana mass), for different assumptions for the values of the nuclear matrix elements that enter into the calculation of $m_{\beta\beta}$.}\label{fig:betabetaf}
\end{figure*}

We have also performed a forecast to compute the expected
sensitivities to $m_{\beta\beta}$ from future cosmological
data. Combining CORE and DESI, for a NH scenario, we obtain the
$95\%$~CL bounds of  $m_{\beta\beta} <
0.034\,\eV$ and $0.005\,\eV< m_{\beta\beta}  <0.053\,\eV$, assuming that $M_\nu=0.06$~eV and $M_\nu=0.1\,\eV$, respectively.
Notice that for the $M_\nu=0.1\,\eV$ case the expected limit on
$m_{\beta\beta}$ is very close to the current one, as for this
particular scenario future cosmological measurements will most likely
be unable to determine the neutrino mass ordering. Figure
\ref{fig:betabetaf} depicts the two-dimensional contours in the
$M_\nu-m_{\beta\beta}$ plane for the two possible fiducial models
above mentioned, together with the expected upper bounds (assuming $m_{\beta\beta} =  0$) from a
future, nEXO-like~\cite{nEXO} neutrinoless double beta decay
experiment. In the case of $M_\nu=0.1\,\eV$
the prospects of observing a positive signal from a future
$0\nu2\beta$ decay are very good (provided neutrinos are
Majorana and the mass mechanism is responsible for $0\nu2\beta$ decay), despite the fact that the hierarchy can not be determined
via cosmological measurements. In this situation the hierarchy
could be extracted by neutrinoless double beta decay itself, since a
positive signal characterized by $m_{\beta\beta} \simeq 0.05\,\eV$ would
suggest an IH scenario. Alternatively, a
positive signal characterized by $m_{\beta\beta} \simeq 0.02\,\eV$ plus the expected sensitivity of $\sigma(m_{\beta\beta})\sim0.01\,\eV$ would point to a NH scenario and discard the IH scenario with high statistical significance.

If, on the other hand, $M_\nu$ is closer to the minimal value allowed in the NH scenario, the sensitivity of future
neutrinoless double beta decay searches may not be enough to
detect the putative signal from Majorana neutrinos, due to the possible disruptive interference played by oscillation parameters in the definition of the Majorana mass. Nevertheless, in that case, future cosmological data can single
out the neutrino mass ordering with high significance. 

\section{Conclusions}\label{sec:conclusion}
We have presented constraints on cosmological parameters in the
context of a $\lcdm+M_\nu$ scenario, with $M_\nu$ representing the
sum of neutrino masses, assuming three massive non-degenerate
eigenstates and properly taking into account the neutrino mass ordering. Indeed, the novelty of the study presented here relies on our
treatment of the neutrino mass ordering, currently totally unknown. We
implement the neutrino hierarchy ambiguity by means of a
\textit{hyperparameter} $\htype$ to be marginalized
over. This approach allows us to \textit{(i)} model the exact mass
splittings without making use of approximations, and including the
information from oscillation measurements; \textit{(ii)} to
quantitatively assess the preference for one of the two hierarchies in
a straightforward fashion, without the need for computing the Bayesian
evidence for performing model comparison, and \textit{(iii)} to account for the incomplete knowledge of the neutrino hierarchy that could potentially affect the neutrino mass bounds. 
We have employed current cosmological data coming fom the Planck satellite measurements of the CMB anisotropies and a compilation of BAO measurements at different redshifts. We have also performed forecasts for future cosmological missions, such as the proposed CMB satellite mission COrE and the future galaxy survey DESI. 

Focusing on current cosmological measurements, we have shown that CMB
temperature and polarization data alone are not sensitive enough to discriminate
between the two hierarchies. When BAO information is included, present
 cosmological probes start to be weakly sensitive to the mass
 ordering ($3:2$ or $4:3$ odds in favor of NH, with or without CMB small-scale polarization, respectively), although compelling
 evidence for one of the two is still lacking. Marginalizing over the
 hierarchy parameter slightly worsens the neutrino mass limits, albeit galaxy
 clustering data in the form of BAO measurements lead to results very
 similar to those obtained in the absence of the  \textit{hyperparameter} $\htype$.
 
Concerning future experiments, their combination turns out to be really powerful, as
it will lead to a $9:1$ preference for the normal hierarchy scenario
versus the inverted hierarchy one, assuming a fiducial cosmology with
a sum of neutrino masses $M_\nu=0.06\,\eV$. However, for larger masses
$M_\nu=0.1\,\eV$ distinguishing the hierarchy via cosmological
measurements alone turns out to be an extremely difficult task. 
Adding other possible future improvements, as, for instance, a $1\%$
prior in the value of the Hubble constant, will not change significantly the
results, since the COrE mission is expected to provide a smaller uncertainty on
$H_0$. Additional constraining power might come from more precise measurements
of the shape of the matter power spectrum, provided that systematics and uncertainties 
related to the exact modelling of the perturbation behaviour in the non-linear regime (where we expect neutrinos to leave their most peculiar signature on the matter power spectrum) are kept under control.

We have also studied the implications and the complementarity with
neutrinoless double beta decay searches. Current limits from the
KamLAND-Zen experiment are competitive and consistent with the
tightest cosmological limit we find here on the \textit{effective
  Majorana mass}, $m_{\beta\beta} <0.056\,\eV$.  These results imply
that, if nature has chosen the inverted hierarchy scheme and the
Majorana neutrino character (versus the normal hierarchy scenario and
the Dirac nature), a positive signal from neutrino neutrinoless double
beta decay searches, as well as a cosmological  detection of the
neutrino mass, could be imminent. Future prospects for neutrinoless
double beta decay experiment are promising for a total neutrino mass
$M_\nu=0.1\,\eV$, regardless of the neutrino mass hierarchy. However, if the lightest neutrino mass eigenstate turns
out to be zero, and the hierarchy normal, the detection of this
putative signal cannot be guaranteed, even for an ultimate, highly
sensitive neutrinoless double beta decay experiment. The good news is
that if this is the case realized in nature, cosmology will be able to
tell us about the neutrino mass hierarchy with compelling statistical significance.

%

\appendix
\section{Results obtained with a logarithmic prior on $\mlight$}

The results discussed in this work rely on the choice of a uniform prior probability distribution for $\mlight$. In this appendix, we would like to comment about the outcome one could obtain, should a uniform prior probability distribution for $\mathrm{log}(\mlight)$ be chosen\footnote{After this work was completed, a paper from Simpson et al.~\cite{Simpson:2017qvj} appeared, claiming strong evidence in favour of normal hierarchy by combining oscillation data and cosmological bounds on $M_\nu$. In that work, a family of Gaussian priors over $\mathrm{log}(m_i)$ (where $m_i=m_1,m_2,m_3$ are the masses of the three neutrino eigenstates) is assumed, and are subsequently marginalized over. Schwetz et. al (including the authors of this work) have already replied with a Comment \cite{Schwetz:2017fey} to Ref.~\cite{Simpson:2017qvj}.}.

\begin{figure}
\begin{center}
\includegraphics[width=\columnwidth]{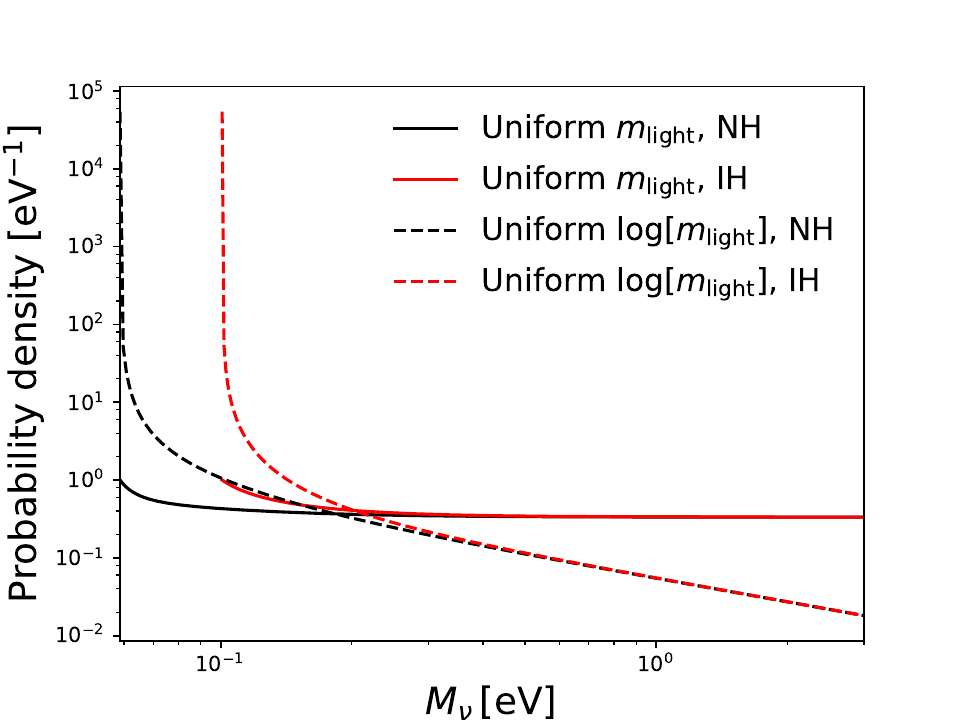}
\end{center}\caption{Prior probability distribution of the sum of the neutrino masses $M_\nu$ in the case of uniform prior over $\mlight$ (solid) and $\mathrm{log}(\mlight)$ (dashed) when assuming either normal (black) or inverted (red) hierarchical mass scenarios.}
\label{fig:pMnu}
\end{figure}

The posterior probability for either NH or IH, as defined in Eq.~(\ref{eq:pnh}), can be written as (in the formulas we focus on NH for the sake of conciseness)
\begin{equation}\label{eq:pNHofMnu}
\mathcal{P}_\mathrm{NH} \equiv \int dM_{\nu}\, \mathcal{L}(\vec{d}\, | M_{\nu},\, \htype=\mathrm{NH}) \, \Pi(M_{\nu}, \htype=\mathrm{NH}).
\end{equation}
It is of course completely equivalent to write the integral in terms of $M_{\nu}$ or $\mlight$; however, the former choice allows for a simplification,
since we know that, to a good approximation, the likelihood of cosmological data does not depend directly on the neutrino hierarchy. Moreover, 
the joint prior probablity is
\begin{equation}\label{eq:priorMnu}
\Pi(M_{\nu}, \htype=\mathrm{NH})=\Pi(M_{\nu}\, | \, \htype=\mathrm{NH})\Pi(\htype=\mathrm{NH}) \, ,
\end{equation}
$\Pi(M_{\nu}\, | \, \htype=\mathrm{NH})$ being the prior probability of $M_{\nu}$ subjected to the choice of the hierarchy, and $\Pi(\htype=\mathrm{NH})$ the prior probability of the NH. Then we can recast Eq. \ref{eq:pNHofMnu} as
\begin{align}\label{eq:pNHofMnu2}
&\mathcal{P}_\mathrm{NH} \equiv \Pi(\htype=\mathrm{NH}) \times \nonumber\\
&\times \int dM_{\nu}\, \mathcal{L}(\vec{d}\, | M_{\nu}) \, \Pi(M_{\nu} | \htype=\mathrm{NH}).
\end{align}
We always assume equal prior probability for the two hierarchies throughout this work, i.e. $\Pi(\htype=\mathrm{NH})=\Pi(\htype=\mathrm{IH})=0.5$. 
It is straightforward to obtain the prior probability of $M_{\nu}$ (conditioned by the hierarchy) that appears in the integral on the right-hand side of Eq.~(\ref{eq:pNHofMnu2}) from the prior probability distribution of $\mlight$. The prior probabilities for $M_\nu$ are shown in Fig.~\ref{fig:pMnu} in the two cases of uniform prior on $\mlight$ (solid) and $\mathrm{log}(\mlight)$ (dashed), given normal (black) and inverted hierarchy (red). The normalisation of the distributions follows from the normalisation of the original distribution of $\mlight$ and $\mathrm{log}(\mlight)$. In the case of a flat prior for $\mathrm{log}(\mlight)$, a lower cutoff has to be imposed and we choose 
${\mlight}>10^{-4}$. In both cases, the upper cutoff is chosen to be $\mlight=1\,\eV$.

A uniform sampling over $\mlight$ is mapped into an almost uniform pior distribution of $M_{\nu}$, with a slight preference for values of the sum of the neutrino masses close to the minimal mass allowed in each hierarchical scenario. Nevertheless, in the limit of very low masses, the distribution never diverges.
On the other hand, a uniform sampling over $\mathrm{log}(\mlight)$ is mapped into a divergent distribution of $M_{\nu}$ for values of the total neutrino mass close to the minimal mass in each hierarchical scenario, with large masses being suppressed as $1/(M_{\nu}-M_{\nu,\mathrm{min}})$. Only the lower cutoff that we have imposed on $\mlight$ prevents the distribution to formally diverge for $M_\nu \to M_{\nu,\mathrm{min}}$.

Regardless of the choice of the sampling over $\mlight$, the distributions for the two hierarchies merge for large values of the masses, as they should in the degenerate regime.

The last piece of information we need in order to compute probability odds for the hierarchies is the likelihood for cosmological data. For the sake of semplicity, at first order we can approximate $\mathcal{L}(\vec{d}\, | M_{\nu})$ as a Gaussian distribution in $M_{\nu}$, centered in the fiducial value for $M_{\nu}$ ($\bar{M}_{\nu}$), with standard deviation given by the sensitivity of the dataset under scrutiny ($\sigma_{M_\nu}$). In this appendix, we present the results for the following cases: current data (Planck TT,TE,EE+lowP and BAO) with $\bar{M}_{\nu}=0\,\eV$ and $\sigma_{M_\nu}=0.08\,\eV$; future data (COrE+DESI) with $\bar{M}_{\nu}=0.06\,\eV$ or $0.1\,\eV$ and $\sigma_{M_\nu}=0.02\,\eV$. 

\begin{figure}
\begin{center}
\includegraphics[width=\columnwidth]{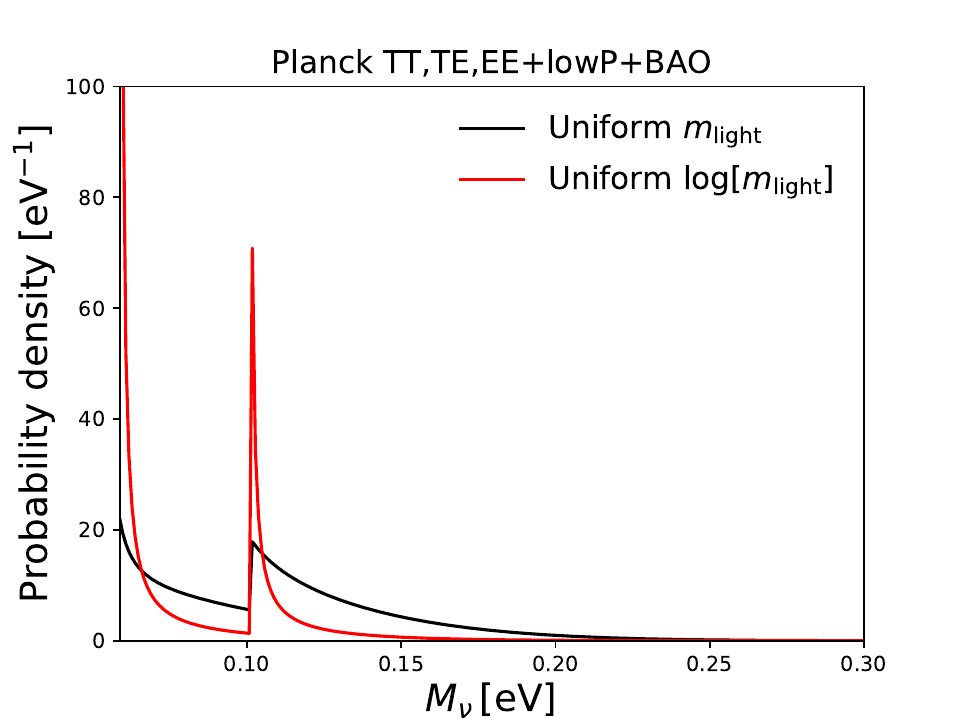}\\\includegraphics[width=\columnwidth]{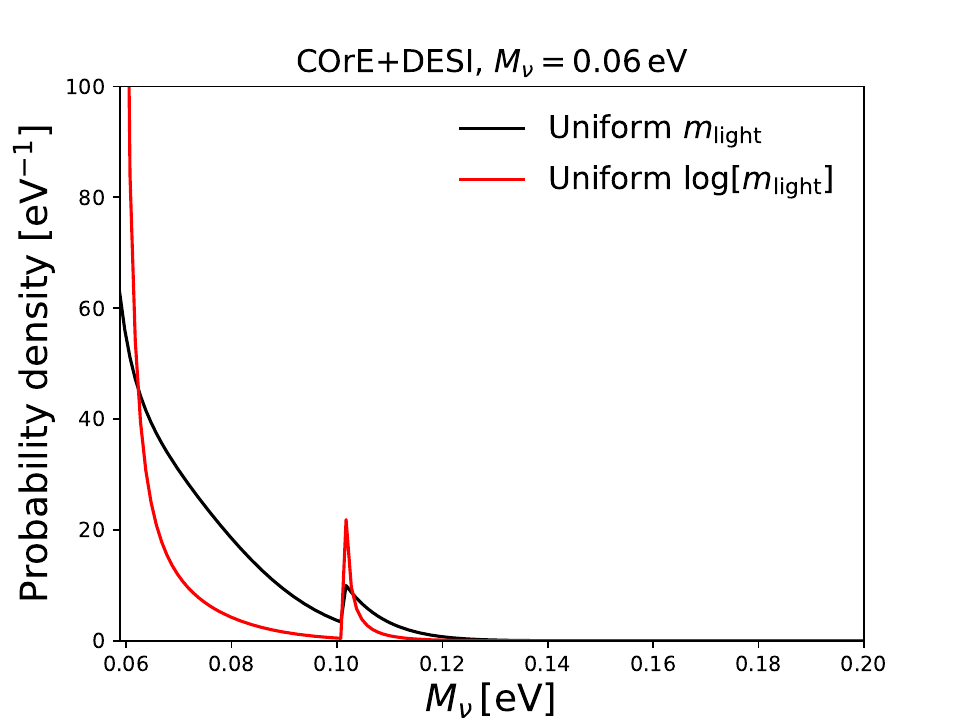}\\\includegraphics[width=\columnwidth]{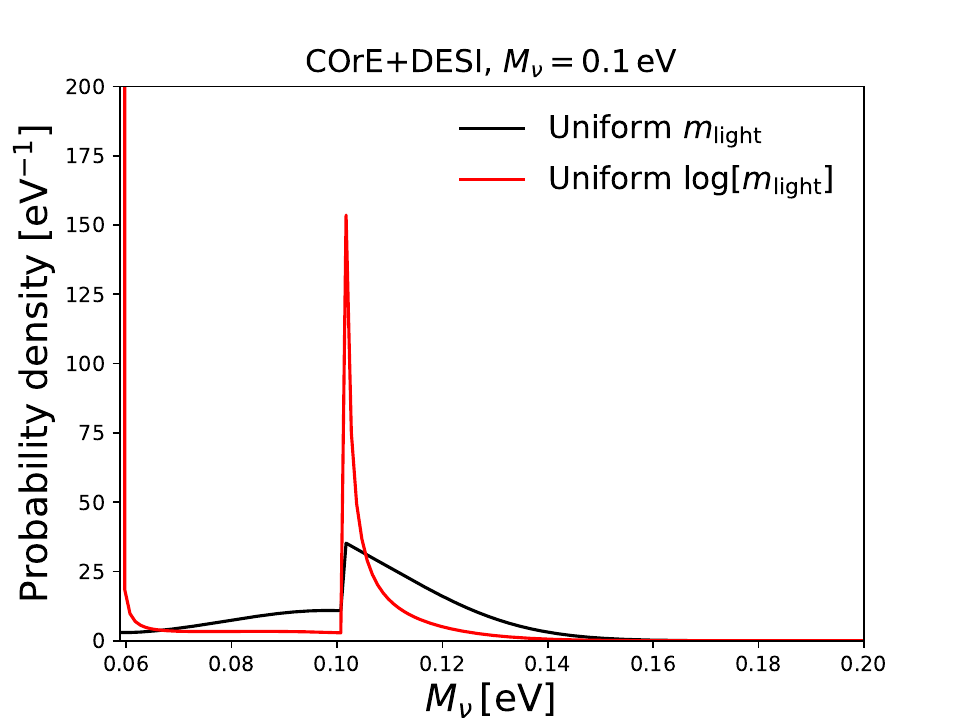}
\end{center}\caption{Posterior probability distribution of the sum of the neutrino masses $M_\nu$ in the case of uniform prior over $\mlight$ (black) and $\mathrm{log}(\mlight)$ (red).}
\label{fig:mnu}
\end{figure}

When integrating Eq.\ref{eq:pNHofMnu} for the different combinations listed above, we find the probability odds of NH versus IH reported in Tab.\ref{tab:odds}. The posterior probability distribution of $M_\nu$ in the different scenarios are reported in Fig. \ref{fig:mnu}.

\begin{table*}
\begin{center}\footnotesize
\scalebox{1.04}{\begin{tabular}{lccc}
\hline \hline
         &Uniform $\mlight$&  Uniform $\mathrm{log}(\mlight)$\\
\hline
\hspace{1mm}\\
Planck TT,TE,EE+lowP+BAO&	$9:5$&	$17:10$\\
\hspace{1mm}\\
COrE+DESI: $M_\nu$= 0.06 eV&	$17:1$&	$10:1$\\
\hspace{1mm}\\
COrE+DESI: $M_\nu$= 0.1 eV&	$6:5$&	$1:3$\\
\hline
\hspace{1mm}\\
\hline
\hline
\end{tabular}}
\caption{Probability odds of normal hierarchy versus inverted hierarchy for the combinations of data reported in the table, for the two different choices of prior probability distribution of $\mlight$.} 
\label{tab:odds}
\end{center}
\end{table*}

The results in the case of a uniform distribution of $\mlight$ are perfectly in agreement with those discussed in the main text and obtained with the full Monte Carlo analysis, both in terms of probability odds and the overall shape of the posterior distribution of $M_\nu$ (compare for example the black curves in Fig.\ref{fig:mnu} with the black dashed curve in the top left panel in Fig.\ref{fig:current} and the dashed curves in the top left panel of Fig.\ref{fig:futuref}). This is a good sanity check and reinforce our confidence in the results obtained in this simple toy model in the case of uniform sampling of $\mathrm{log}(\mlight)$. In this case, we still find a preference for normal hierarchy for the combinations Planck TT,TE,EE+lowP+BAO and COrE+DESI with $M_\nu=0.06\,\eV$, even though milder than in the case of uniform sampling over $\mlight$. Furthermore, in the case of COrE+DESI with $M_\nu=0.1\,\eV$, the uniform sampling over $\mathrm{log}(\mlight)$ results in a mild preference for inverted hierarchy. 

In each case, Fig.\ref{fig:mnu} clearly shows that the posterior distributions of $M_\nu$, marginalized over the hierarchy, are highly squeezed towards the minimal mass allowed in each hierarchical scenario when we sample uniformly over $\mathrm{log}(\mlight)$, as expected given the shape of the prior distributions depicted in Fig.\ref{fig:pMnu}.

We also checked that we are able to reproduce these results by performing a full Monte Carlo analysis. Even in the case in which we assume a fiducial model with $M_\nu=0.1\,\eV$ distributed according to the normal hierarchy scenario, the final result is a mild preference for inverted hierarchy.

As expected, we find that the results are strongly driven by the prior choice, given that we assumed that the likelihood for cosmological data is independent of the hierarchy and only depends on the sum of the neutrino masses (i.e. to the total energy density in neutrinos). We argue that this is a fair approximation, since even if there are well known physical effects induced by a different hierarchical distribution of the masses among the three eigenstates on the cosmological probes, these effects are well out of the sensitivity reach of current and future cosmological experiments. 
This means that the choice of the prior should be addressed and motivated carefully and that the role of the prior should be always made clear when discussing final results. 

In our work, we decide to employ a uniform distribution for $\mlight$ since we wanted to use a parameter which could fit different datasets and not just cosmology. In other words, while it is true that cosmological datasets are at first order sensitive to the sum of neutrino masses instead of the mass of the single eigenstates, the same is not true for laboratory experiments, such as kinematic measurements from beta decay, or searches for neutrinoless double beta decay. We wanted to be as general as possible and leave open the possibility to incorporate external datasets in our analysis.

Furthermore, as it is clear from Fig.\ref{fig:pMnu}, the choice of a uniform distribution of $\mlight$ reflects in an almost flat distribution of $M_\nu$. We think this is a good choice, as it allows us to implicitly employ a (nearly) uniform distribution over the parameter that cosmological data are more directly sensitive to, namely $M_\nu$. On the contrary, a uniform distribution of $\mathrm{log}(\mlight)$ highly favours values of $M_\nu$ closer to the minimal mass allowed in each hierarchical scenario. We argue that, in a situation when the likelihood is not informative enough, the choice of a uniform distribution for $\mlight$ better represents our ignorance about the value of this parameter.

 \section*{Acknowledgements}
We would like to thank Allen Caldwell, Alex Hall, Steen Hannestad and Thomas Schwetz-Mangold for useful comments and discussion.
Based on observations obtained with Planck (http://www.esa.int/Planck), an ESA science mission with instruments and contributions directly funded by ESA Member States, NASA, and Canada. We also acknowledge use of the Planck Legacy Archive.
M.G. and K.F. are supported by the Vetenskapsr\aa det (Swedish Research
Council, Contract No. 638-2013-8993). M.L. acknowledges support from ASI through ASI/INAF Agreement 2014- 024-R.1 for the Planck LFI Activity of Phase E2. O.M. is supported by PROMETEO II/2014/050, by the Spanish
Grant FPA2014--57816-P of the MINECO, by the MINECO Grant
SEV-2014-0398 and by the European Union’s Horizon 2020
research and innovation programme under the Marie Skłodowska-Curie grant
agreements 690575 and 674896. K.F. acknowledges support by the DoE grant DE-SC0007859.

%

\end{document}